\documentclass{elsart1p}
 \usepackage{graphicx}
\usepackage{graphicx}
\usepackage{amssymb}
\def\lsim{\raise0.3ex\hbox{$<$\kern-0.75em\raise-1.1ex\hbox{$\sim$}}}
\def\gsim{\raise0.3ex\hbox{$>$\kern-0.75em\raise-1.1ex\hbox{$\sim$}}}

\begin{document}
\begin{frontmatter}
%
%
%
%
%
\title{Equation of State from Lattice QCD Calculations}
%
%

\author{Rajan Gupta}

\address{Theoretical Division, Los Alamos National Lab, Los Alamos, N.M. 87545, USA}

\begin{abstract}
We provide a status report on the calculation of the Equation of State
(EoS) of QCD at finite temperature using lattice QCD.  Most of the
discussion will focus on comparison of recent results obtained by the
HotQCD and Wuppertal-Budapest (W-B) collaborations.  We will show that
very significant progress has been made towards obtaining high
precision results over the temperature range of $T=150-700$ MeV. The
various sources of systematic uncertainties will be discussed and the
differences between the two calculations highlighted.  Our final
conclusion is that the lattice results of EoS are getting precise
enough to justify being used in the phenomenological analysis of heavy
ion experiments at RHIC and LHC.
\end{abstract}

\end{frontmatter}

\section{The Road to Precision Lattice QCD Calculations}
\label{sec:Intro}

One of the goals of simulations of lattice QCD is to provide a precise
non-perturbative determination of the EoS of QCD
over the temperature range $150-700$ MeV that is being probed in
experiments at RHIC and the LHC.  The EoS, along with the transition
temperature $T_c$ and transport coefficients such as shear viscosity,
are crucial inputs into phenomenological hydrodynamical models used to
describe the evolution of the quark gluon plasma (QGP).  In this talk
I will mainly review the two recent and most complete calculations of
the EoS by the HotQCD~\cite{HoTQCDeos:2009} and
Wuppertal-Budapest~\cite{WBeos:2010} collaborations.

Simulations of lattice QCD at finite temperature are carried out on a
4-D hypercube of size $aN_\tau \times aN_S \times aN_S \times aN_S $
where $a$ is the lattice spacing usually denoted in units of
GeV${}^{-1}$ or fermi. The spatial size $aN_S$ is taken large enough
so that finite volume corrections are under control and small. Past
calculations show that for finite temperature simulations with
$N_S/N_\tau = 4$ these corrections are smaller than statistical errors
for $T \lsim 3T_c$. For higher temperatures larger $N_S$ may be
required. In QCD, the gauge coupling $\beta\equiv 6/g^2$ is related to
$a$ by dimensional transmutation and the continuum theory is recovered
in the limit $a \to 0$ or equivalently $g \to 0$ or $\beta \to
\infty$. To provide a perspective on how fine the current lattice
simulations are note that $N_\tau=10$ corresponds to $a \approx 0.1$
fermi or $\sim 2$ GeV${}^{-1}$ at the transition temperature $T \sim
200$ MeV.

The second set of control parameters (inputs) in the simulations are
the quark masses. The results discussed in this review are for $2+1$
flavors, $i.e.$, two flavors of degenerate $u$ and $d$ quarks and a
heavier strange quark. In nature $2 m_s / (m_u + m_d) \approx 27.5$
and $\overline{m} = (m_u + m_d)/2 \sim 3.5$ MeV and $m_s \sim 90$
MeV. The small value of $\overline{m}$ provides the key computational
challenge because the most time consuming part of the simulations is
the inversion of the Dirac operator, a very large sparse matrix. This
inversion is done using interative Krilov solvers that have critical
slowing down in the limit $ \overline{m} \to 0$. The computational
cost increases as $m^{-7/2}$ or faster and calculations become very
expensive with decreasing quark mass. For this reason $2+1$ flavor
simulations are typically done fixing the strange quark mass to its
physical value and simulating at a number of values of
$\overline{m}/m_s$ from which extrapolations to the physical
$\overline{m}$ are made.  Recent calculations by the W-B and HotQCD
collaborations show that the computer power has reached a stage where
simulations can be done close to, or directly at, physical
$\overline{m}$. Thus, in the state-of-the-art simulations, this source
of systematic error (extrapolation in $\overline{m}$) is now under
good control.

The tuning of the set of parameters $\{g, m_s, \overline{m}\}$ is done
as follows.  One first fixes the ratio $ m_s / \overline{m}$, ideally
to $ m_s / \overline{m} = 27.5$. Then for a judiciously chosen value
of $g$, zero temperature simulations are done to measure two
independent physical quantities whose values are experimentally
measured or well determined, and one of which is sensitive to strange
quark mass, for example the K-meson mass $M_K$ and the pion decay
constant $f_\pi$ (or $f_K$ and $M_\pi$). The value of $m_s$ is tuned
until the lattice results for these two quantities match their
physical values.  This fixes $a$ and $m_s$.  Now depending on how
finely one wants to scan in $T$ (or $a$) a new value of $g$ is chosen
and the value of $m_s$ is again tuned to reproduce the observables,
thus determining the new $a$ keeping $m_s / \overline{m}$ fixed.  This
process generates a set of $\{g, m_s, \overline{m}\}$ values for
which, by construction, the physics (defined by matching lattice $M_K$
and $f_\pi$ to physical values) is fixed. This line in the $\{g, m_s,
\overline{m}\}$ space, since $m_s$ is tuned to the physical value and
$ m_s / \overline{m}$ is fixed, is called a line of constant physics
(LoCP).  The utility of simulating along LoCP is to reduce the three
dimensional space of input parameters to a line along which only the
lattice spacing is changing. This procedure provides better control
over taking the continuum limit.  The extent to which $\{g, m_s\}$
would have varied had one chosen two different physical quantities,
say $M_{\overline{s} s}$ and $M_N$, is a measure of variations in
discretization errors along different LoCP. The emphasis of the
lattice community is to present results extrapolated to $a \to 0$ in
order to remove these discretization errors, which are proportional
to powers of $a$, by simulating at a number of values of $a$. In case
of finite temperature calculations we extrapolate results at fixed $T$
to $N_\tau \to \infty$ by simulating at a number of values of
$N_\tau$.

The lattice size in the Euclidean time direction defines the
temperature of the system by the relation $T=1/aN_\tau$. The scale $a$
for fixed $\{g, m_s, \overline{m}\}$ is the same for zero-temperature
and finite temperature lattices. Thus, knowing $a$ corresponding to a
given $g$ and quark masses uniquely determines $T$ for a given
$N_\tau$. An important consequence of the fact that $g$ or
equivalently $a$ or $T$ is the single parameter that controls lattice
simulations is that only one thermodynamic quantity can be determined,
which for the extraction of EoS is the trace anomaly $I/T^4 \equiv
(\varepsilon - 3 p)/T^4$, where $I$ is called the integration measure.

The above approach for scanning in $T$ is called the fixed $N_\tau$
approach.  A second approach that I will not cover in detail and which
is being pursued by the WHOT collaboration~\cite{WHOTeos:2010} with
improved Wilson fermions is called the fixed $a$ approach.  In this
approach, for a given $a$ (the same process is used for fixing $\{g, m_s,
\overline{m}\}$) one simulates on a number of different $N_\tau$
lattices to scan in $T$.  The advantage of this approach is that only
a single zero-temperature matching calculation, needed to carry out
subtractions of lattice artifacts in finite $T$ data, is required for
each $a$.  The weakness of this approach is that the scan in $T$ is
limited by the coarseness and range of $N_\tau$ values possible,
$i.e.$ $N_\tau = 6, 8, 10, 12, 14$, before one runs out of
computational power. The recent results using the fixed $a$ approach
by the WHOT collaboration~\cite{WHOTeos:2010} are very encouraging and
I refer interested readers to their paper for details.

\section{Taste Symmetry breaking with Staggerd fermions}
\label{sec:Taste}

In the naive discretization of the Dirac action there automatically is
a $2^4$ doubling of flavors. In the staggered fermion approach, using
a lattice symmetry called ``spin diagonalization'', the degeneracy is
reduced from 16 to 4 by placing a single degree of freedom at each
lattice site. Under this construction a $2^4$ hypercube is the basic
unit cell that reduces to a point in the continuum limit. The 16
degrees of freedom in the cell represent, in the continuum limit, four
identical copies (called taste) of Dirac fermions. On the lattice this
four-fold degeneracy gives rise to a proliferation of particles
propagating in the QCD vacuum, for example, there are 16 pions
distinguished by their taste rather than just one.  If taste symmetry was
unbroken at finite $a$, the 4-fold degeneracy could be handled by just
dividing results by the appropriate degeneracy factor.  Problems arise
because this degeneracy is broken at finite $a$ and one does not know
$a\ priori$ how large this systematic error is and how it effects
simulations of QCD thermodynamics in particular. The most common
approach to quantify this effect is to study the difference in the
masses of the 16 pions, and determine how these differences vanish
when lattice results are extrapolated to the continuum limit.

\begin{figure}
\includegraphics[width=0.5\hsize,angle=0]{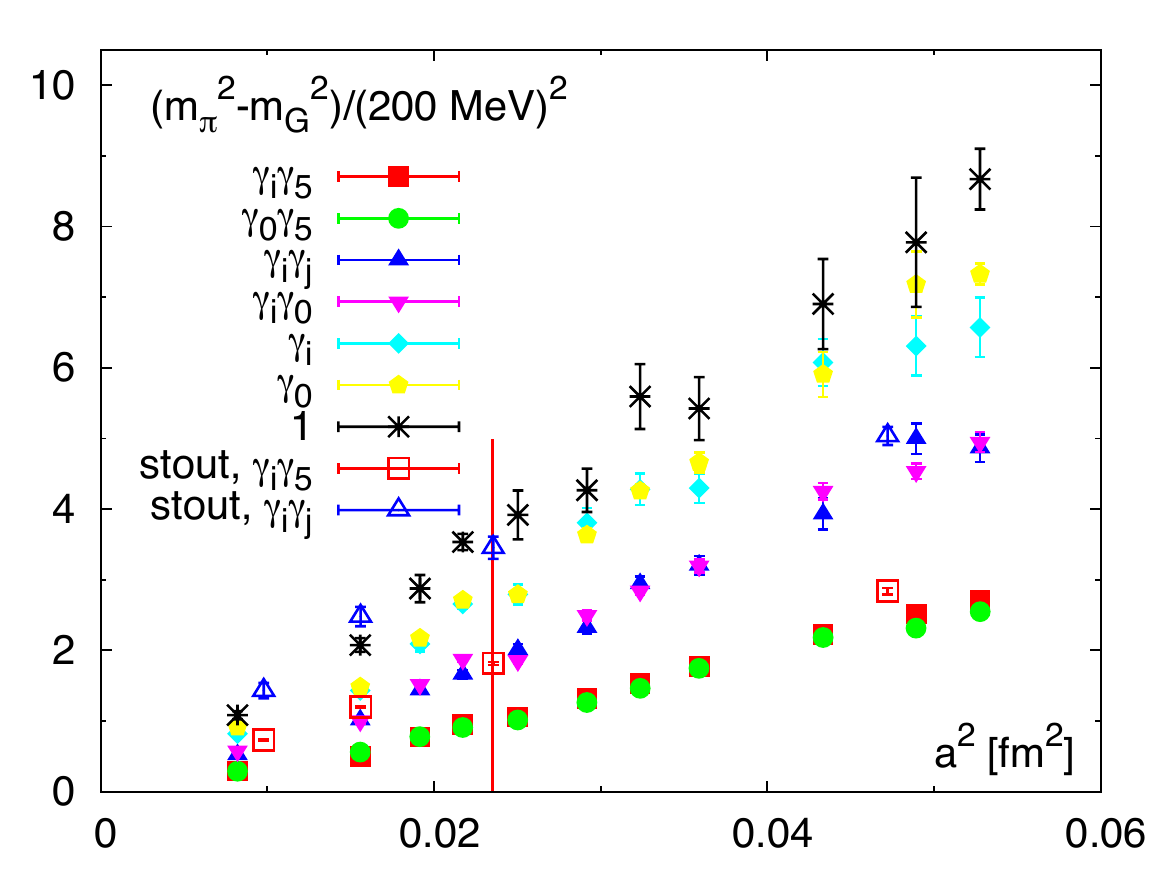}%
\includegraphics[width=0.5\hsize,angle=0]{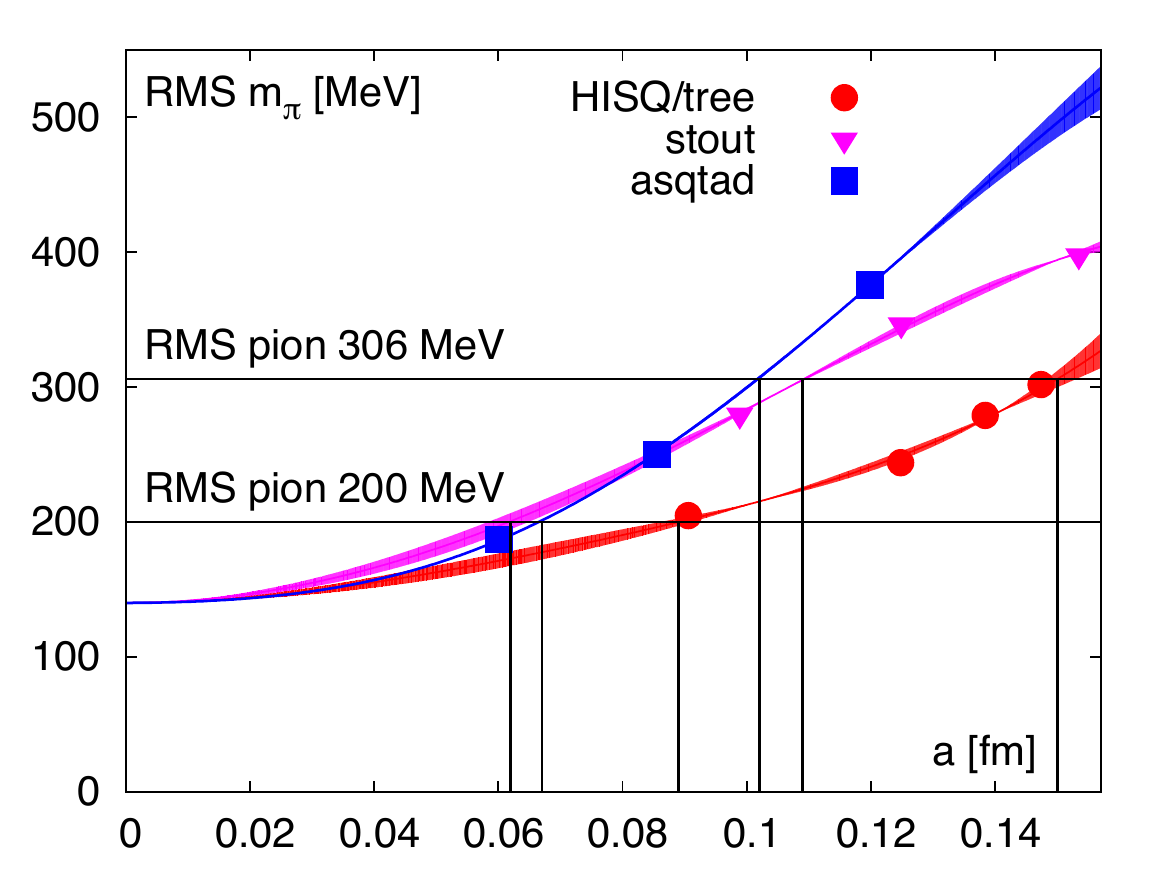}%
\caption{The splitting between the 16 pions due to taste symmetry
  breaking for HISQ/tree, stout and asqtad actions (left panel). The right
  panel shows the root mean squared mass corresponding to a Goldstone
  pion mass of $140$ MeV. Note that the HISQ action used so far for
  thermodynamics does not include tadpole improvement of the gauge
  action and is therefore called HISQ/tree in~\cite{HoThisq:2010}.}
\label{BP_pion_splits_hisq}
\end{figure}

The HotQCD collaboration has studied the consequences of taste
symmetry breaking utilizing three versions of improved staggered
fermions $-$ the asqtad, p4 and HISQ/tree
formulations~\cite{HoThisq:2010}.  In Fig.~\ref{BP_pion_splits_hisq}
we show preliminary HotQCD HISQ/tree results for $M^2_\pi - M^2_G$
where $M_G$ is the Goldstone pion mass versus $a$, and also
compare with results from the stout and asqtad actions. The large
spread in masses that increases on coarser lattices shows that taste
symmetry is indeed badly broken~\cite{Follana:2006}. Based on such
studies the conclusion is that at any given $a$, the taste breaking is
least in HISQ/tree followed by stout, asqtad, and p4 actions. One
consequence of this discretization error for thermodynamics is that
the contribution of any state, for example the pion, is not just from
the lowest taste state (the Goldstone pion) but is some weighted
average of the 16 pions (or the appropriate multiplets for other states).
Thus, at any given $a$ the effective masses of all hadrons are larger
than the desired ground state value. The magnitude of the effect is shown in
the right panel of Fig.~\ref{BP_pion_splits_hisq} which plots the root
mean squared mass of the 16 pion states corresponding to a Goldstone
pion mass of $140$ MeV. Results at low temperatures, $T < 150$ MeV,
and on small $N_\tau$ lattices are most susceptible to this
discretization error.  Taste breaking also puts a caveat on the above
described tuning of quark masses: setting $ m_s / \overline{m}=27.5$
does not guarantee that simulations have been done at the physical
light quark masses. Simulating at a number of $ m_s / \overline{m}$
values to control the chiral extrapolation followed by taking the
continuum limit using a number of $N_\tau$ lattices provides the best
understanding of systematic errors and for obtaining physical results.

A second issue with staggered simulations that include the strange quark (or 
in future the charm quark) is the need to take the
fourth root of the determinant to compensate for the 4-fold
degeneracy. Creutz~\cite{Creutz:2007} claims that
this ``rooting'' is a fundamental flaw of the staggered formulation,
however, its effect in current simulations may be small since the
strange quark mass is large, whereas the review by Sharpe~\cite{Sharpe:2006} (covering a
large body of work) shows that while the staggered formulation may be
ugly, it gives physical results in the continuum limit. (For
degenerate $u$ and $d$ quarks, this rooting problem is overcome
because of a lattice symmetry whereby the square root of the
determinant can be written as the determinant on even (or odd) sites.)
In a perfect world one would like to use a Wilson-like action that
maintains the continuum flavor structure and chiral symmetry, $i.e.$,
domain wall or overlap fermions, however, to date most
thermodynamical simulations use staggered fermions for two reason:
they are much faster ($10-20\times$) to simulate than even simple Wilson
fermions and because of a residual chiral symmetry that protects the
Goldstone pion.

\section{The Trace Anomaly}
\label{sec:Trace}

\begin{figure}
\includegraphics[width=0.5\hsize,angle=0]{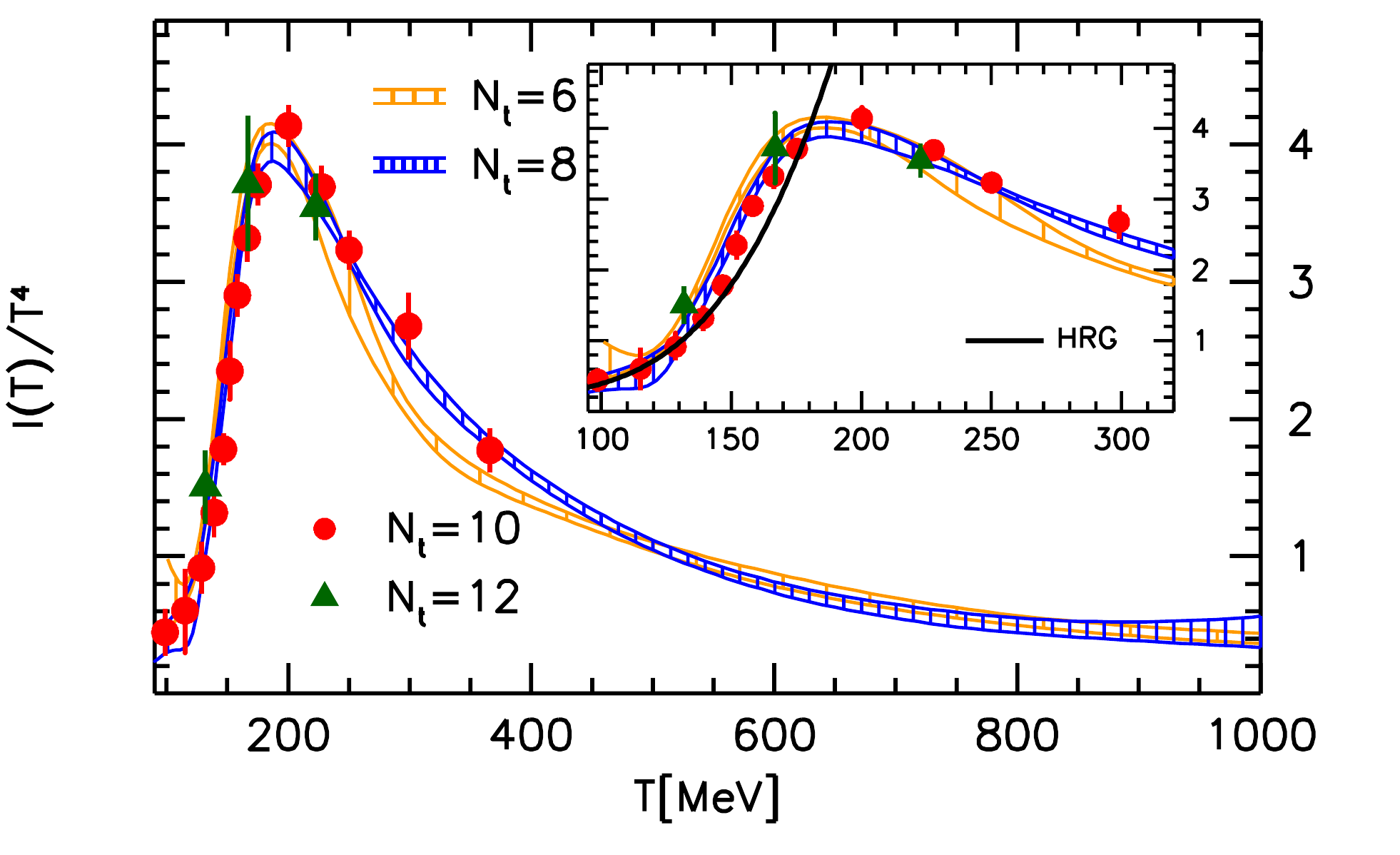}%
\includegraphics[width=0.5\hsize,angle=0]{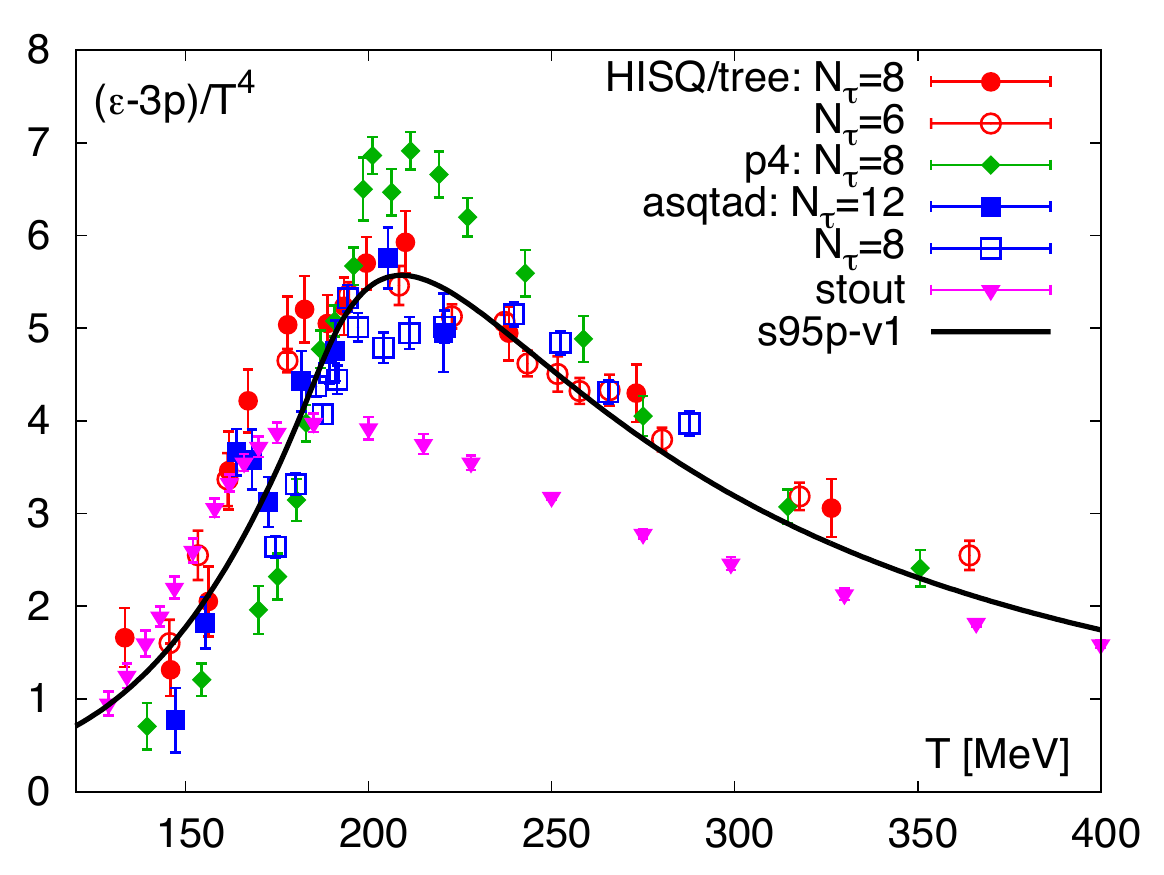}%
\caption{Integration measure $I/T^4$ from W-B (left) and preliminary
  HotQCD (right) comparing different actions. The curve (right panel)
  is a a parametrization of $\varepsilon-3p/T^4$ that is based on HRG and the
  p4/asqtad lattice data at high temperatures~\cite{Huovinen:2010};
  and the W-B data are shown in purple for comparison.}
\label{fig:WBeos_I}
\end{figure}

The results from the HotQCD~\cite{HoTQCDeos:2009}~\cite{HoThisq:2010}
and W-B~\cite{WBeos:2010} collaborations for $I/T^4
\equiv (\varepsilon - 3 p)/T^4$, the single thermodynamic quantity
calculated on the lattice, are shown in Fig.~\ref{fig:WBeos_I}.

Before making detailed comparisons it is important to stress here, and
applicable to all discussion that follows, that the HotQCD results do
not yet incorporate extrapolation to the physical quark mass or the
continuum limit. The most extensive data are for $N_\tau=6,8$ and $
m_s /\overline{m} =10$ with new ongoing calculations at $ m_s
/\overline{m}=20$ and $N_\tau=12$. The W-B results at $N_\tau=6,8$
have been obtained for number of values of $ m_s /\overline{m}$
including at $ m_s /\overline{m} = 28.15$ where the final physical
value is quoted. (Recall, however, the caveat about the uncertainty in
locating the physical value of $\overline{m}$ due to taste breaking.)
The W-B data at $N_\tau=10$ and $12$ are more limited in $T$ values
and are at $ m_s /\overline{m}=28.15$ only.  W-B define their
$N_\tau=8$ results to represent the continuum value. Data with $N_\tau=10$ at
$T \leq 365$ MeV (red points in left panel of Fig.~\ref{fig:WBeos_I})
and the three points (green points in Fig.~\ref{fig:WBeos_I}) with
$N_\tau=12$ provide a consistency check since they show
no significant discretization effects relative to $N_\tau=8$
data. 

The overall form of the results by the two collaborations is
similar. There are, however, two significant differences between the
HotQCD and W-B data. The first is the value of $I$ at the peak, $\sim
5.8$ versus $4.1$, and the peak in the W-B data is shifted to lower
$T$ by about $20$ MeV. W-B collaboration attribute these differences
to the lack of extrapolation of HotQCD data in quark mass and $a$,
$i.e.$, residual discretization errors. Preliminary HotQCD results
with $N_\tau=8$ lattices using HISQ/tree fermions (also shown in
Fig.~\ref{fig:WBeos_I}) give a similar peak height and position as the
asqtad action (the p4 results are higher but decreasing with
$N_\tau$). The agreement between asqtad and HISQ/tree and since the
HISQ/tree action is the more improved than stout, has smaller
discretization errors and less taste breaking, it is not clear if,
today, we have a simple resolution of the difference. The forthcoming
results with HISQ/tree action on $N_\tau=8$ and $12$ lattices being
simulated by the HotQCD collaboration should help clarify these
issues.

\section{Pressure, Energy Density, Entropy and Speed of Sound}
\label{}
\begin{figure}
\includegraphics[width=0.5\hsize,angle=0]{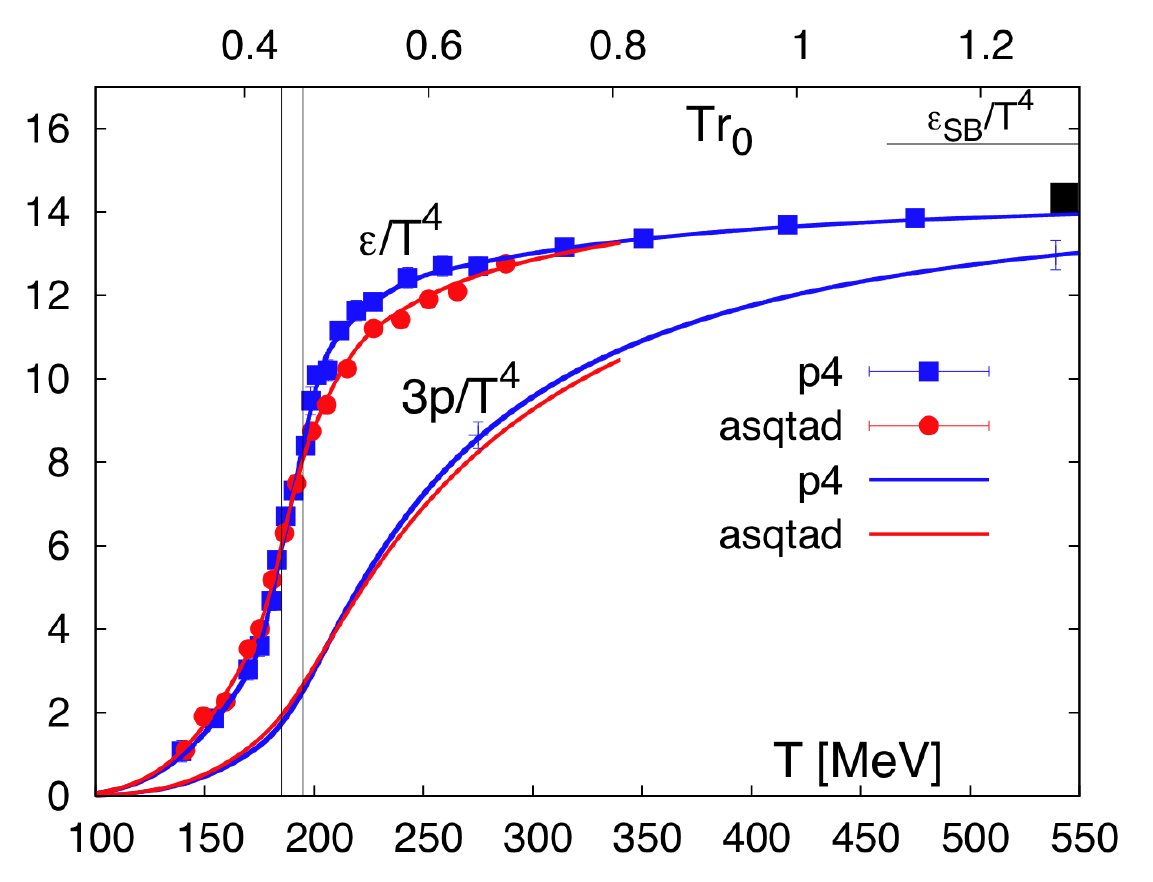}
\caption{HotQCD collaboration~\cite{HoTQCDeos:2009} results for
  $\varepsilon/T^4$ and $3p/T^4$ on $N_\tau=8$ lattices with
  $m_s/\overline{m}=10$.}
\label{fig:HoTe_and_3p}
\end{figure}

\begin{figure}
\includegraphics[width=0.5\hsize,angle=0]{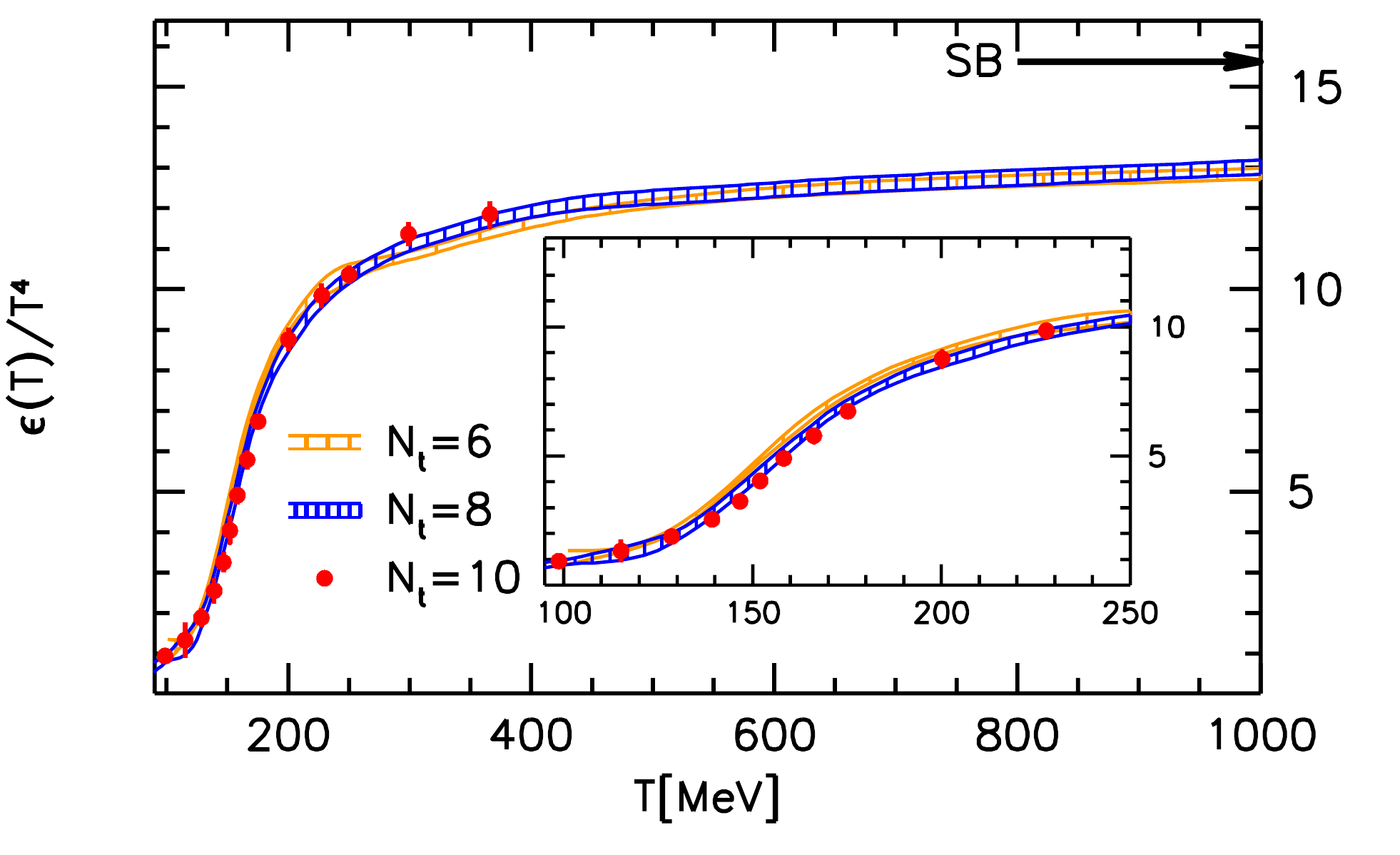}%
\includegraphics[width=0.5\hsize,angle=0]{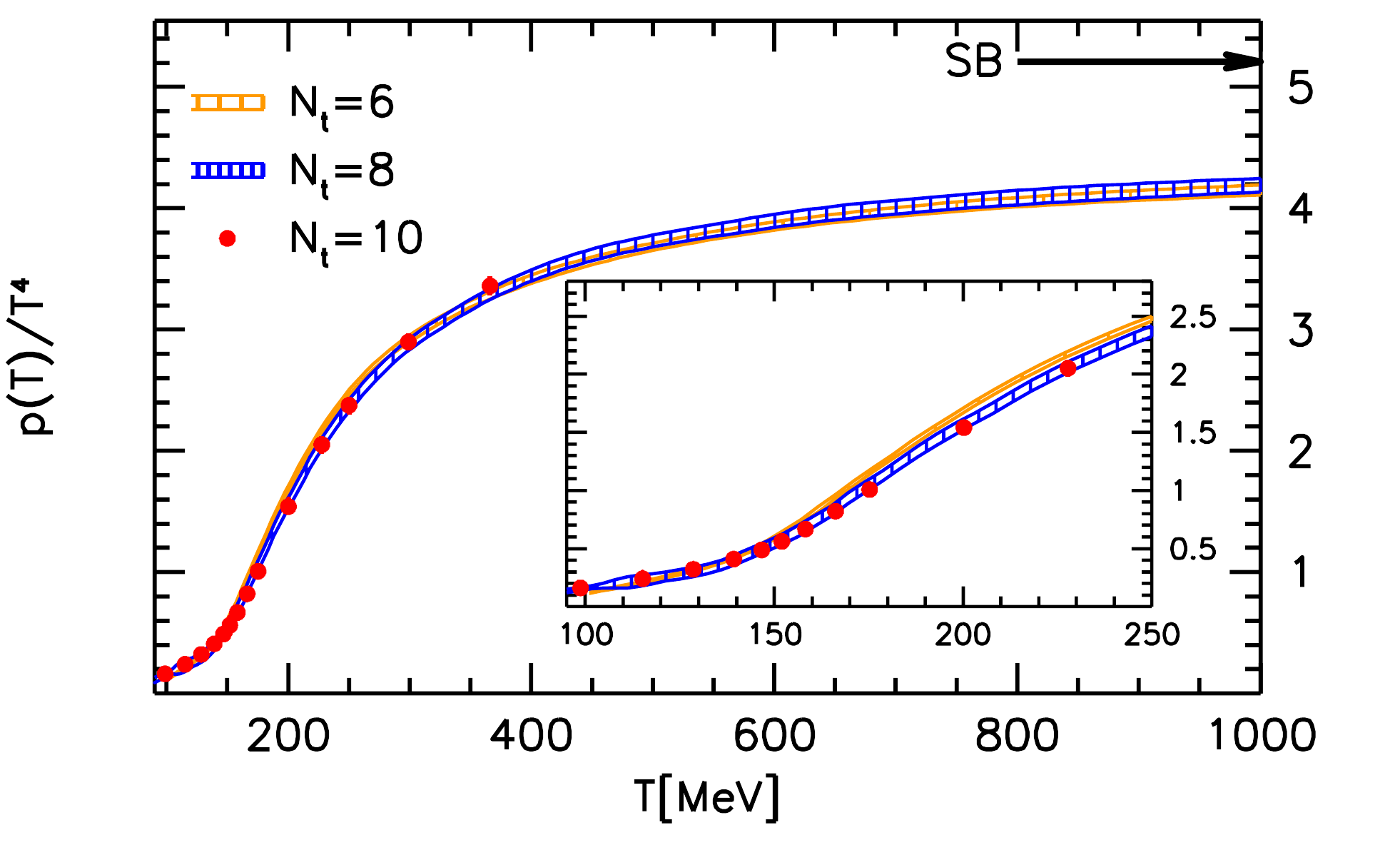}
\caption{Results for $\varepsilon/T^4$ and $p/T^4$ from the W-B
  collaboration~\cite{WBeos:2010}.  To compare results for $p$ note
  that the W-B data are for $p/T^4$ whereas the HotQCD results in
  Fig~\ref{fig:HoTe_and_3p} are for $3p/T^4$.}
\label{fig:WBeos_e}
\end{figure}

The pressure $p$ 
can be determined from the trace anomaly using the following relations:
\begin{eqnarray}
\frac{I}{T^4} \equiv \frac{\Theta^{\mu\mu}(T)}{T^4}  &\;& \equiv  
\frac{\varepsilon - 3 p}{T^4} = T\frac{\partial}{\partial T} \left( \frac{p}{T^4}\right) \,
\label{eq:I}\\
\frac{p(T)}{T^4} - \frac{p(T_0)}{T_0^4} &\;& =  \int_{T_0}^T dt \frac{\Theta^{\mu\mu}(t)}{t^5}  
\label{eq:p}
\end{eqnarray}
The results for pressure and energy density are summarized in
Figs.~\ref{fig:HoTe_and_3p} and~\ref{fig:WBeos_e}.  To obtain pressure
$p$ there are two issues that need to be addressed when carrying out
the integration in Eq.~\ref{eq:p}. The first is to construct a smooth
function that represents the lattice data for $(\varepsilon - 3
p)/T^4$ over the whole range of $T$ since $I/T^4$ has been calculated
only at a finite number of values of $T$. The second is the choice of
$T_0$ above which $\Theta^{\mu\mu}(T_0)/T_0^4$ is well-determined and
at which point $p(T_0)$ can be estimated reliably.

The HotQCD collaboration has investigated a number of ansatz for
parameterizing $\Theta^{\mu\mu}(T_0)/T_0^4$ and find that the results
for $p$ do not vary significantly.  The uncertainty due to the ansatz
is shown by the error bars on $p/T^4$ at $T=275,\ 540$ MeV in
Fig.~\ref{fig:HoTe_and_3p}.  The W-B collaboration uses a variant of
the method $-$ they parameterize the pressure itself and then evaluate
its derivatives to match to $I$. The band in Fig~\ref{fig:WBeos_e} show the uncertainty.

The second issue is more significant.  The systematic errors in
lattice data grow as $T$ is lowered and are expected to be large below
$T=150$ MeV. At the same time $p(T=150)$ is not negligible and
$a\ priori$ unknown.  One approach is to use the hadron resonance gas
(HRG) model for $p(T=150\ {\rm MeV})$. This requires that there be
reasonable agreement between the HRG and lattice values at $T=150$
MeV. The HotQCD data approaches the HGR from below and at $T=150$ MeV
there is a significant difference. Another approach is to use
$T_0=100$ MeV where there is more confident in the HRG value but then
one has to confront the uncertainty in matching and parameterizing
$\Theta^{\mu\mu}(T_0)/T_0^4$ between $T=100-150$ MeV.  The HotQCD
collaboration use $p=0$ at $T=100$ MeV for their central value and the
HRG value to estimate the uncertainty whose magnitude is shown by the
black square on $\varepsilon/T^4$ at $T=550$ MeV in
Fig.~\ref{fig:HoTe_and_3p}.  The W-B collaboration show that a
modified ``lattice'' HRG calculation, taking into account taste
breaking in pion and kaon states, fits the lattice data between
$T=100-140$ MeV.  Nevertheless, they choose $p(T=100MeV,
m_s=\overline{m}) = 0$ for the normalization.

\begin{figure}
\includegraphics[width=0.5\hsize,angle=0]{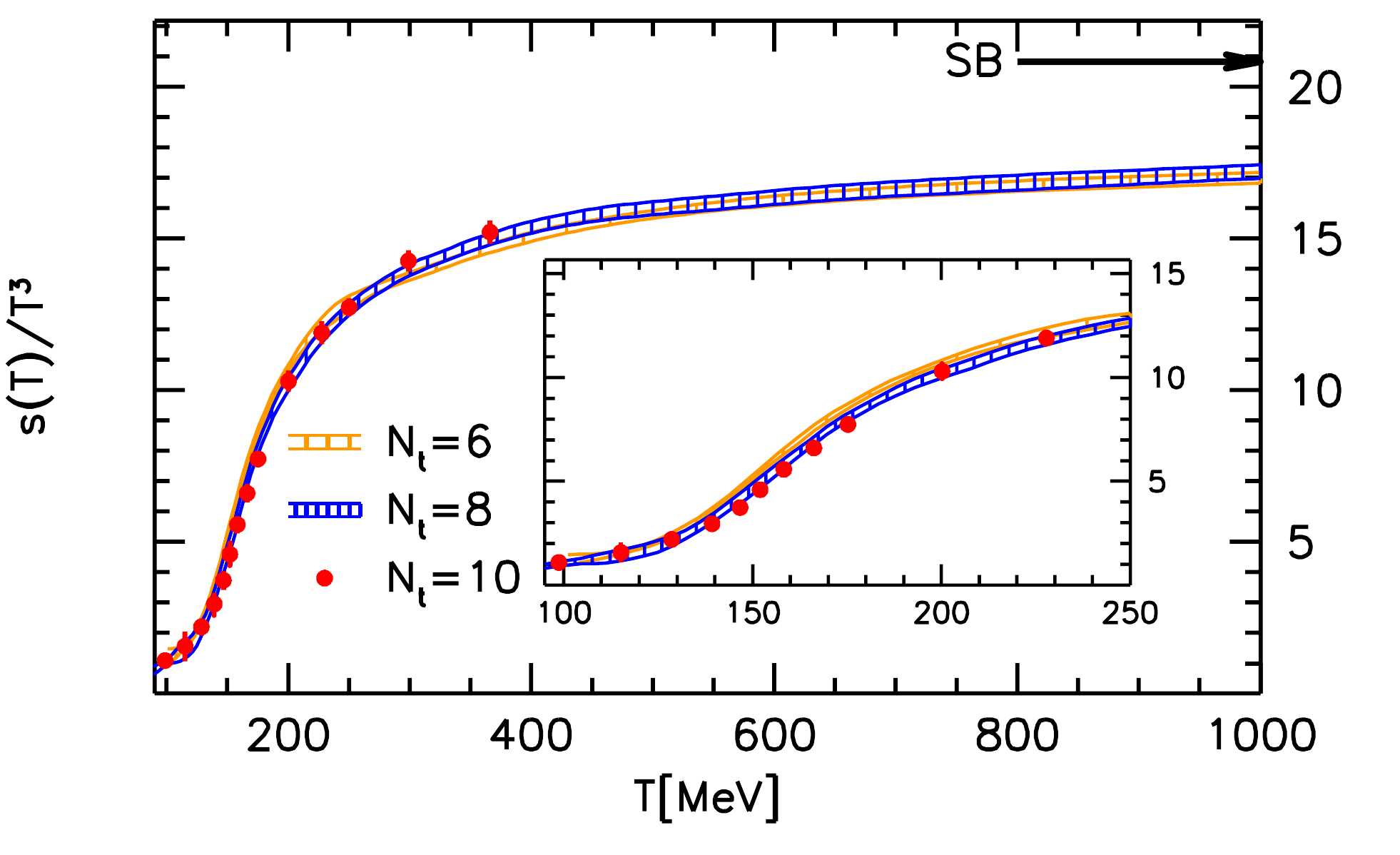}%
\includegraphics[width=0.5\hsize,angle=0]{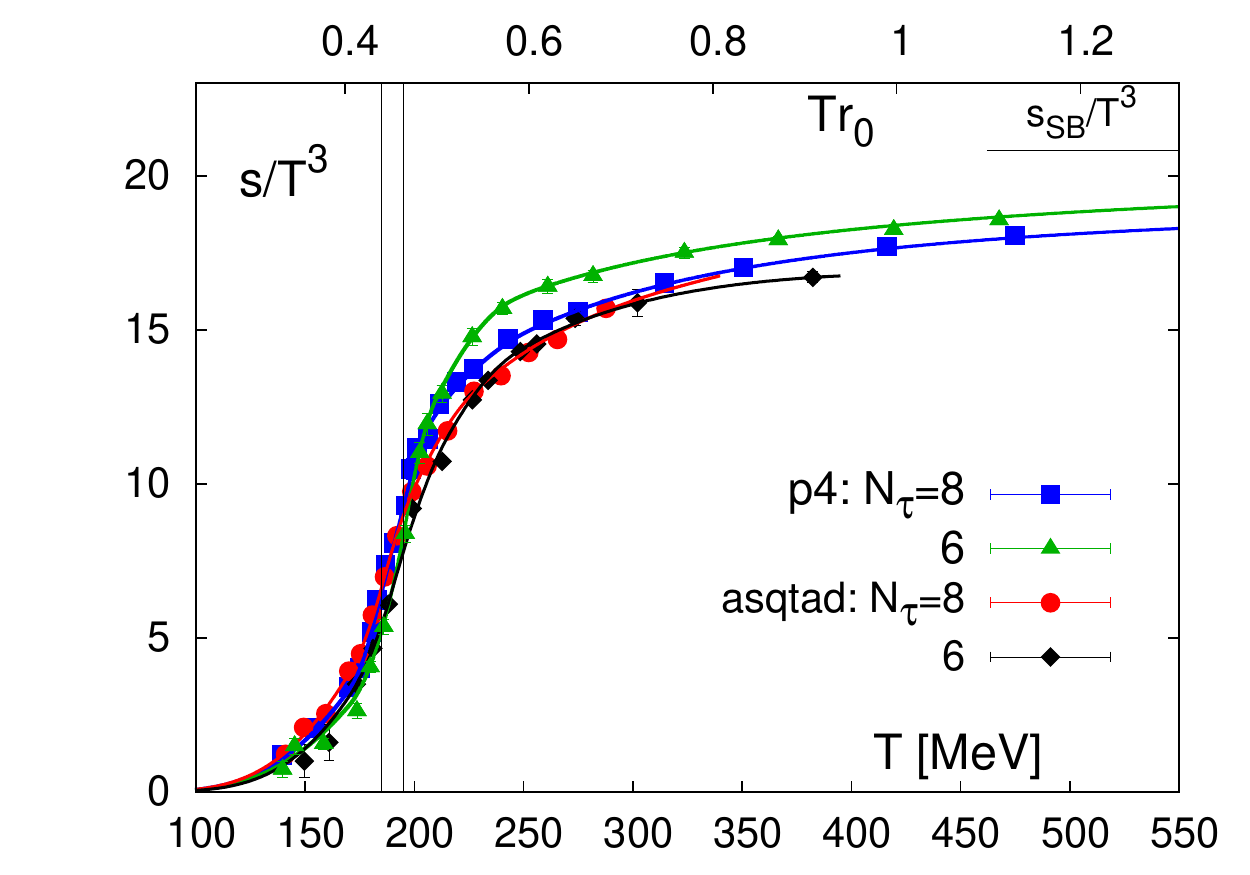}
\caption{Comparison of entropy density obtained by W-B (left) and HotQCD (right) collaborations. 
The HotQCD data are for $m_s/\overline{m}=10$.}
\label{fig:WBeos_s}
\end{figure}

Once $I$ and $p$ are determined the energy density is given by
$\varepsilon/T^4=I/T^4+3p/T^4$, entropy by $s=(\varepsilon+p)/T$ and
the speed of sound $c_s$ by
\begin{equation}
c_s^2 = \frac{dp}{d\varepsilon} = \varepsilon \frac{d(p/\varepsilon)}{d\varepsilon} + \frac{p}{\varepsilon} \;\; .
\label{eq:cs}
\end{equation}
A comparison of results for $\varepsilon$,  $p$ and $s$ is shown in
Figures~\ref{fig:HoTe_and_3p}, \ref{fig:WBeos_e}, and \ref{fig:WBeos_s}.

The W-B collaboration apply two corrections to the estimate for
$p(T)$. The first is to guarantee that the lattice results for each
$N_\tau$ match the continuum Stefan-Boltzmann value at $T=\infty$. To
do this they construct the ratio of the continuum Stefan-Boltzmann
value for $p$ to its free-field ($T=\infty$) lattice value (the
continuum integrals are replaced by lattice sums appropriate for each
$N_\tau$) and then correct the lattice data at all $T$ by this ratio.
This ratio is large, $1.517$ and $1.283$ for $N_\tau=6$ and $8$
respectively. Since $N_\tau=8$ data are used to define the continuum
estimate, this correction is too large to justify on the basis of
a tree-level improvement of lattice observables (lattice operators
used to probe the physics). Furthermore, this correction is also
applied to $I$, $\varepsilon$ and $s$.  The second correction made by
the W-B collaboration is to shift upward their results for $p/T^4$ by
half the difference, $0.06$, between the lattice and HRG estimates at
$T=100$ MeV. Again, to me, this is not a well-motivated correction of
data.  Future simulations and better understanding of the low $T$
region will hopefully alleviate the need for such corrections.

\begin{figure}
\includegraphics[width=0.5\hsize,angle=0]{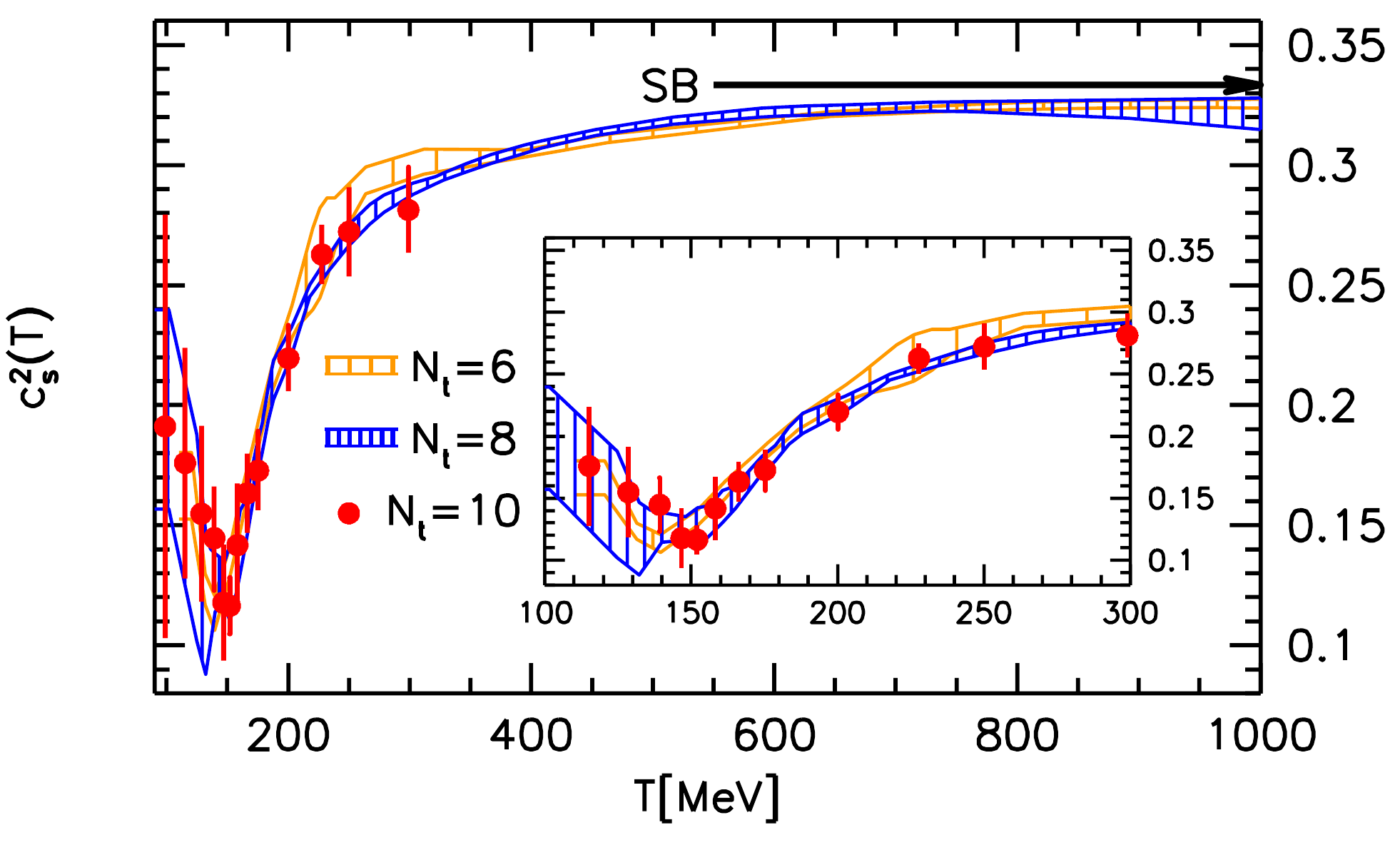}%
\includegraphics[width=0.5\hsize,angle=0]{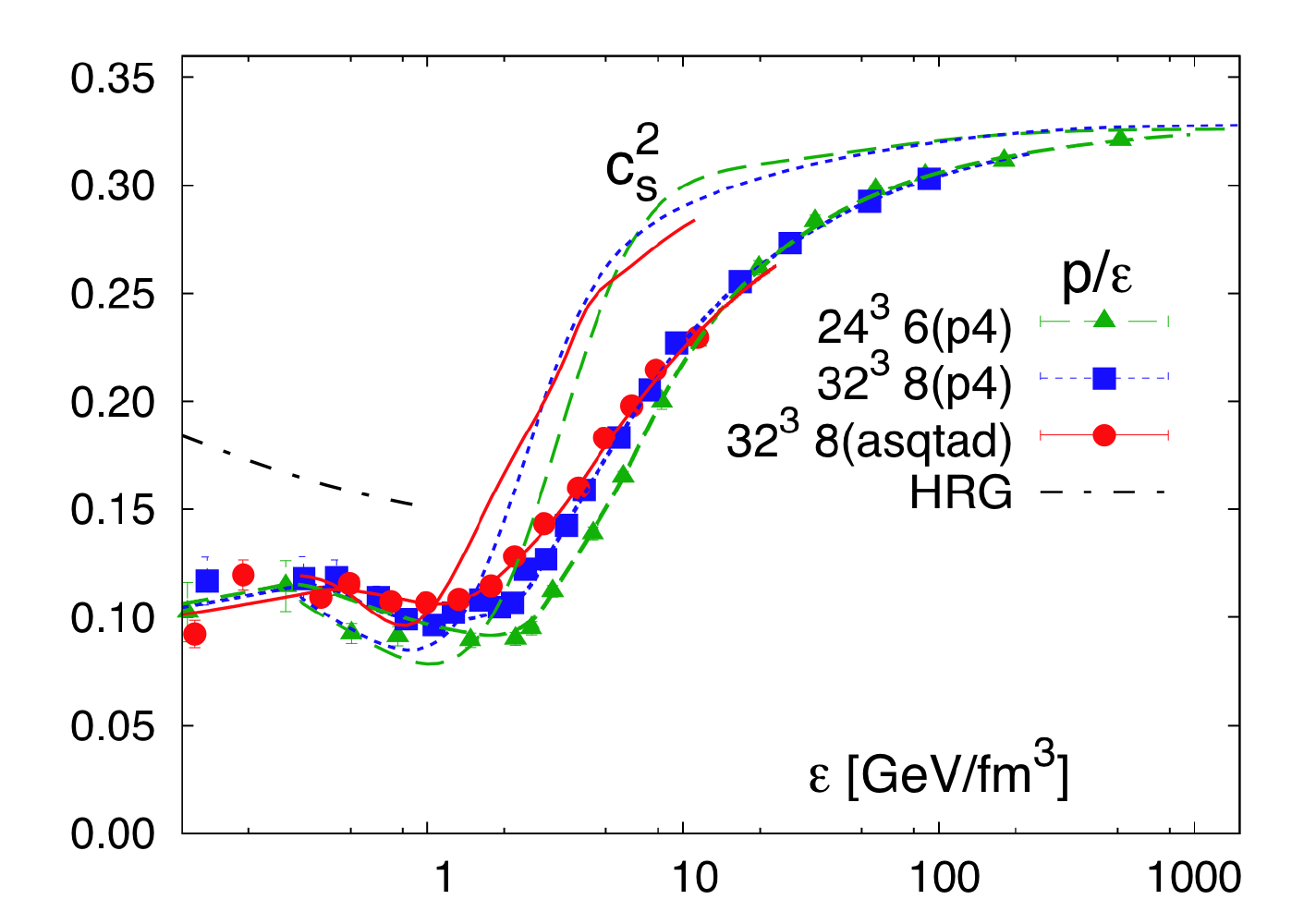}
\vskip -0.2in
\caption{Comparison of the speed of sound $c_s^2$. The W-B data (left
  panel) are plotted versus $T$.  The HotQCD data for $p/\varepsilon$
  and $c_s^2$ versus the energy density $\varepsilon$ using the fits
  to $\varepsilon$ and $p$ (right panel).}
\label{fig:WBeos_cs}
\end{figure}

A comparison of results for the speed of sound are shown in
Fig.~\ref{fig:WBeos_cs}.  The fundamental quantity needed to calculate it is
${p/\varepsilon}$ as shown in Eq.~\ref{eq:cs}. Two features in the data
are worth commenting on. First, data in Fig.~\ref{fig:WBeos_e} show
that, in the transition region from hadronic matter to QGP, the
energy density is changing more rapidly than the pressure. This 
implies that $c_s$ should show a dip in the transition region as is
indeed observed. Second, $c_s$ rises quickly after the transition region and
reaches close to the relativistic Boltzmann gas value of $1/3$ by
$T \sim 400$ MeV.

\vskip -0.1in

\section{Prospects for improvement in the EoS of QCD in the near future}
\label{sec:future}

Significant progress has been made in determining the EoS using
lattice QCD in the last three years. The current lattice results for
the EoS have already given the Heavy Ion community a much better
understanding of the dynamics of the QGP. Lattice estimates are now
being used in hydrodynamic analysis of the evolution of the QGP in
experiments at the RHIC at BNL and at the LHC.

There are a number of ways in which both HotQCD and W-B 
collaborations are improving their estimates:
\begin{itemize}
\item
The HotQCD collaboration will present their estimates of the continuum
values with HISQ/tree action on $N_\tau = 6,\ 8$ and $12$ lattices and
with $m_s/\overline{m} = 10$ and $20$.
\item
Both collaborations will include the charm quark to provide results
with ($2+1+1$) dynamical flavors.  Preliminary partially-quenched
estimates suggest that the charm quark contribution starts to become
large at above $T \sim 300$ MeV.
\item
To fully control finite volume effects at $T>500$, both collaborations
will simulate on larger spatial volumes, larger than $N_s/N_\tau = 4$.
\item
In addition to simulations with staggered fermions, simulations with
improved Wilson and domain wall fermions are
maturing~\cite{WHOTeos:2010} \cite{Cheng:2009} and will provide
independent checks of the staggered results in, hopefully, the near
future.
\end{itemize}

With these improvements a number of unresolved issues such as the
location and height of the peak in $I$, control over systematic errors
at low temperatures, and the impact of charm quark at high
temperatures, should be addressed over the next couple of years. So
stay tuned.

\vskip 0.1in
{\bf Acknowledgements:}  
I thank T. Nayak, R. Verma and P. Ghosh for the invitation to
a very informative conference and acknowledge the
support of DOE grant KA140102.

\end{document}